\documentclass[twocolumn,showpacs,preprintnumbers,amsmath,amssymb,groupedaddress,prl]{revtex4}

\usepackage{graphicx}
\usepackage{dcolumn}
\usepackage{bm}
\usepackage{epsfig}

\def\e{\begin{equation}}
\def\f{\end{equation}}
\def\=#1{\overline{\overline #1}}

\def\-#1{{\bf #1}}
\def\.{\cdot}

\begin{document}

\title{Sub-wavelength imaging by wire media}

\author{Pavel A. Belov}
\affiliation{Queen Mary College, University of London, Mile End
Road, London, E1 4NS, United Kingdom}

\author{Yang Hao}
\affiliation{Queen Mary College, University of London, Mile End
Road, London, E1 4NS, United Kingdom}

\author{Sunil Sudhakaran}
\affiliation{Queen Mary College, University of London, Mile End
Road, London, E1 4NS, United Kingdom}

\begin{abstract}
Original realization of a lens capable to transmit images with
sub-wavelength resolution is proposed. The lens is formed by
parallel conducting wires and effectively operates as a telegraph:
it captures image at the front interface and the transmit it to the
back interface without distortion. This regime of operation is
called canalization and is inherent in flat lenses formed by
electromagnetic crystals. The theoretical estimations are supported
by numerical simulations and experimental verification.
Sub-wavelength resolution of $\lambda/15$ and 18\% bandwidth of
operation are demonstrated at gigahertz frequencies. The proposed
lens is capable to transport sub-wavelength images without
distortion to nearly unlimited distances since the influence of
losses to the lens operation is negligibly small.
\end{abstract}


\maketitle

Resolution of common imaging systems is restricted by the so-called
diffraction limit, since they operate only with propagating spatial
harmonics emitted by the source. Conventional lenses can not
transmit evanescent harmonics which carry sub-wavelength
information, since these waves exhibit exponential decay in usual
naturally occurring materials. In order to overcome the diffraction
limit it is required to use the other sort of materials for the
construction of lenses. It is required to engineer an artificial
material (metamaterial) with electromagnetic properties which
dramatically differ from those of materials available in nature
\cite{Smithreview}. One of the options was suggested by Pendry in
his seminal paper \cite{Pendrylens}. Pendry proposed to use
left-handed materials, isotropic media with both negative
permittivity and permeability \cite{Veselago}. The planar slab of
such a metamaterial provides unique opportunities to restore and
even amplify amplitudes of evanescent modes. This becomes possible
due to resonant excitation of surface plasmons at the interfaces of
left-handed material slab. However, the promising theoretical
predictions meets numerous practical difficulties in the creation of
left-handed metamaterials. On the one hand, the major problem is the
creation of materials possessing magnetic properties at optical and
tetrahertz frequencies \cite{YenTHZ,LindenTHZ}. On the other hand,
the issues related to losses play a very important role as well. The
enumerated problems are closely related with fundamental
restrictions and can be hardly overcome. The other option to reach
the sub-wavelength resolution was suggested by Wiltshire
\cite{Wiltshire1,Wiltshire2}. The idea is based on the use of an
array of magnetic wires, so-called Swiss rolls \cite{PendryMagnet}
in order to transfer sub-wavelength information directly from the
source to the image plane (pixel-to-pixel imaging principle). The
lens formed by Swiss rolls have to be placed into the near field of
the source since it is capable to transfer evanescent harmonics
rather than amplify them. This realization of sub-wavelength imaging
experiences similar problems as the design of left-handed medium: it
is required to obtain a metematerial with magnetic properties and
also keep the losses to be small.

\begin{figure}[ht]
\centering \epsfig{file=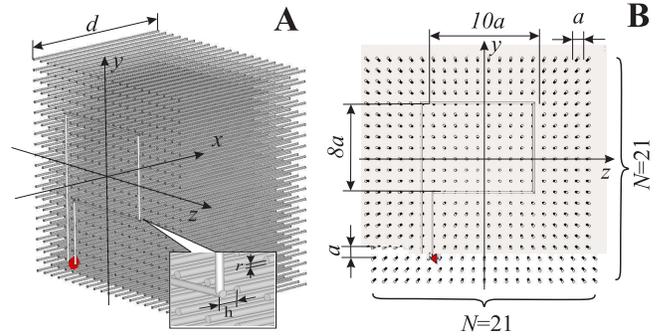, width=8.5cm} \caption{Geometry of
the flat lens formed by wire medium and the source of the form of
letter P. (A) Perspective view. (B) Front view. Parameters:
$a=10$mm, $r=1$mm, $d=150$mm, $h=5$mm, $f=1$GHz.} \label{geom}
\end{figure}

In the present paper we suggest an alternative opportunity of
sub-wavelength lens construction. It does not require magnetic
properties to be created. The imaging device is formed by an array
of parallel conducting wires, so-called wire medium
\cite{Rotmanps,Brown,pendryw,WMPRB}, see Fig. \ref{geom}. At the
first sight it seems that this structure is electrical analogue of
Wiltshire's system. An array of Swiss rolls, being similar to
magnetic wires, is capable to transmit s-polarized (transverse
electric, TE) spatial harmonics of the source spectrum. An array of
wires operates in the same manner, but for p-polarized (transverse
magnetic, TM) waves.
In the other words, an array of Swiss rolls restores at the back
interface normal component of magnetic field produced by the source.
An array of wires restores normal component of electric field. At
the same time, there is a serious difference between the Wiltshire's
system and slab of wire medium. Swiss rolls are artificial resonant
structures which behave as magnetic wires only at the frequencies in
vicinity of the resonance. This fact preconditions Swiss rolls to be
narrow-band and very lossy. The conducting wires in this sense are
natural electrical wires. It means that they are wide-band and
practically lossless. The absence of strong losses (inherent in
Swiss rolls) in ordinary wires lifts restriction on the lens
thickness. It allows to create sub-wavelength lenses of nearly
arbitrary thickness and deliver images with sub-wavelength
resolution into far-field region of the source and beyond. The
imaging system effectively works as a telegraph formed by
multi-conductor transmission line.

Different spatial harmonics incident to the front interface of the
lens formed by a wire medium experience different
reflection/transmission properties. It happens due to impedance
mismatch between air and wire medium. The interface of wire medium
has a surface impedance for p-polarization which is independent on
incidence angle in contrast to the air which surface impedance
varies for different angles of incidence. The reflections from the
thin slabs are negligibly small, but become significant for thick
layers. This problem can be solved by choosing an appropriate
thickness of the slab in order to fulfill condition for Fabry-Perot
resonance and reduce reflections. Actually, the reflections can be
nearly cancelled out at all in the present case in contrast to the
classical case of dielectric slab where it is possible to avoid
reflection for normal incidence, but for oblique incidences some
nonzero reflections are inevitable. For any incidence angle the wire
medium supports propagating modes which travel across the slab with
the same phase velocity equal to the speed of light, so-called
transmission line modes \cite{WMPRB}. If the slab thickness is
chosen to be integer number of half-wavelengths then Fabry-Perot
condition holds for any incidence angle (including complex ones) and
such a slab experiences total transmission effect. This phenomenon
of collective reduction of reflections for all incidence angles
together with the fact that the waves are allowed to transfer energy
only across the slab (along the wires) with a fixed phase velocity
is called as canalization regime \cite{canal}. This regime can be
observed in various electromagnetic crystals which possess flat
isofrequency contours at certain frequencies \cite{Chien,Li,Kuo}.
The wire medium is an unique example of electromagnetic crystals
with such properties observed at very long wavelengths as compared
to the period of the crystal. It provides a possibility to reach
nearly unlimited sub-wavelength resolution. Resolution of the lens
formed by wire medium is restricted only by its period which can be
made as small as it is necessary for certain applications. In the
present case the resolution is equal to double period of the
lattice: two different objects can be distinguished if they are
located close to two different wires, but their location within one
elementary cell can not be determined. It means that the wire medium
lens has the best possible resolution among other periodical
structures since Luo have shown in \cite{Subwavelength} that double
period of the lattice is the maximum ultimate limit for resolution
of superlenses formed by electromagnetic crystals.

\begin{figure}[ht]
\centering \epsfig{file=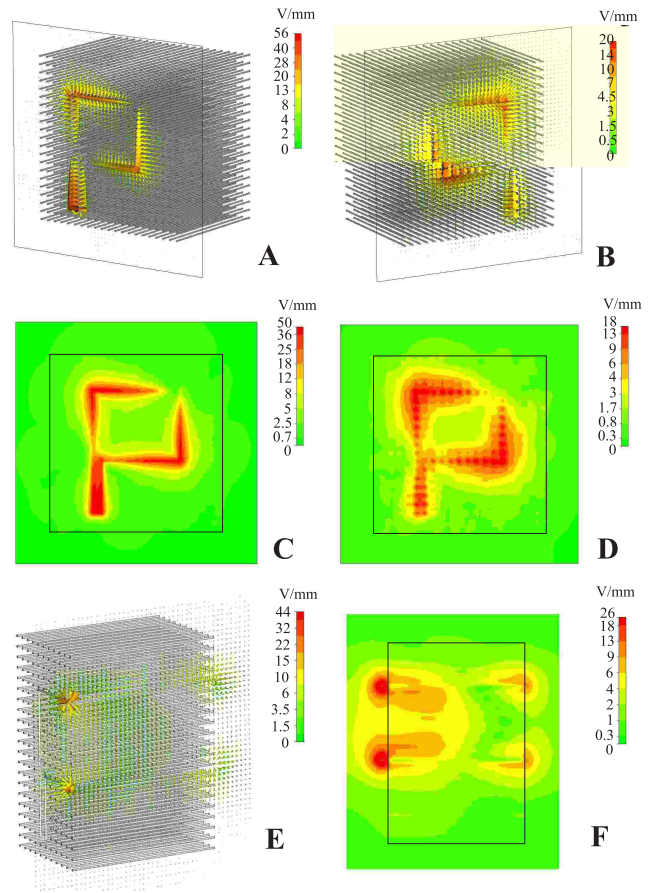, width=8.5cm} \caption{Distribution
of electric field and its absolute value: (A), (C) in vicinity of
source (at 2.5mm distance from the front interface); (B), (D) in
image plane (at 2.5mm distance from the back interface); (E), (F) in
transverse plane} \label{num}
\end{figure}

In order to verify the concept described above, numerical
simulations of the structure presented in Fig. \ref{geom} were
performed using CST Microwave Studio package. The lens consisting of
$21\times 21$ array of aluminium wires excited by a source in the
form of P letter was modelled. The working frequency $f$ is 1GHz,
length of wires (thickness of slab) $d$ is 15cm (half of
wavelength), period of lattice $a$ is 1cm, radius of wires $r$ is
1mm. The source in the form of P letter is placed at $h=5$mm
distance from the front interface of the lens and feeded by a point
current source $I=1$A. Results of the simulation are presented in
Fig. \ref{num}.
The source produces sub-wavelength distribution of electrical field
at the front interface of the slab, see Fig. \ref{num}.A. The
p-polarized contribution of the field is canalized from the front
interface to the back interface and forms an image, see Fig.
\ref{num}.B. The quality of the imaging can be clearly seen in Fig.
\ref{num}.C and D where absolute values of electrical field in
vicinity of the front and back interfaces are plotted. The local
maximums of intensity produced by terminations of the wires are
visible in Fig. \ref{num}.D. The resolution of imaging in the
present case is equal to $2$cm (double period of the structure),
which is one fifteenth of the wavelength ($\lambda/15$).

The canalization principle can be easily illustrated with the help
of Figs. \ref{num}.E,F where distribution of electrical field in
transverse plane is presented. Two sources (visible at the left
sides of Figs. \ref{num}.E,F) produce both s- and p-polarized
spatial harmonics. The s-polarized harmonics practically do not
interact with wire medium since they have electric fields
perpendicular to the thin wires. The lens behaves nearly as an air
for such waves. Therefore, the s-polarized evanescent harmonics
decay with the distance and practically disappear at the center of
the slab: their contribution is visible only in vicinity of the
front interface. In contrast to these waves, the p-polarized
harmonics are guided by wires and canalized from the front interface
to the back one. The trace of their propagation is visible inside
the slab. This waves form an image at the back interface (right
sides of Figs. \ref{num}.E,F).

\begin{figure}[t]
\centering \epsfig{file=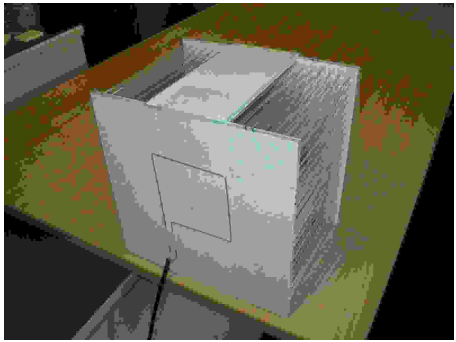, width=6cm} \caption{Photo of wire
medium lens used in the experiment} \label{setup}
\end{figure}
\begin{figure}[t]
\centering \epsfig{file=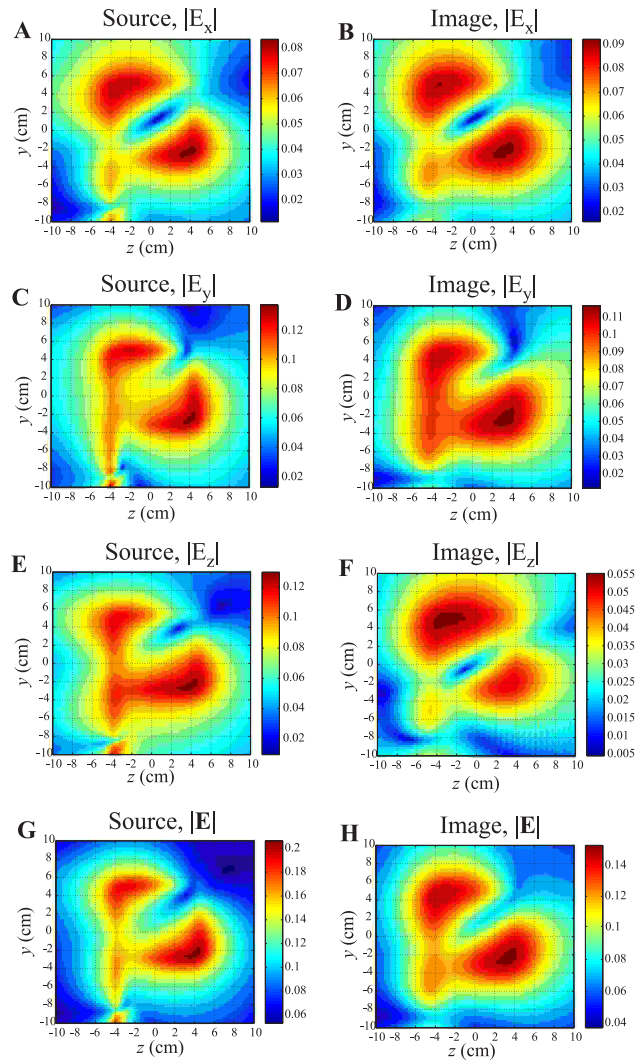, width=8.3cm} \caption{Results of
near field scan measurements: absolute values of x-, y- and z-
electrical field components and total field in arbitrary units:
(A,C,E,G) in source plane and (B,D,F,H) in image plane,
respectively.} \label{m98}
\end{figure}

Excellent correspondence between theoretical predictions and
numerical simulations proves validity of the canalization regime.
Further confirmation is also achieved by experimental verification.
The slab of wire media with parameters as in Fig. \ref{geom} was
constructed. Photo of the lens is available in Fig. \ref{setup}. The
wires are fixed by thin slabs of foam which has relative
permittivity close to unity at frequencies around 1GHz. The source
is an antenna in the form of the P letter directly fed by a coaxial
cable and located at the surface of the foam (approximately 3mm away
from the terminations of wires). The mechanical near-field scanning
device was used for measurements of electrical field distribution in
the source plane (5-7mm from the source and front interface of the
lens) and image plane (4-6mm from the back interface of the lens). A
short piece of wire (1cm long) connected to the central conductor of
coaxial cable was used as a probe. The measurements have been done
for x-,y- and z-orientations of the probe (see Fig. \ref{geom} for
orientation of axes). The best quality of imaging was observed at
980MHz. This frequency corresponds to Fabry-Perot resonance and it
is slightly lower than the theoretical value of 1GHz due to the
tolerance of sample construction. The measurement results are
presented in Fig. \ref{m98}. The x-component of electrical field
(normal to the interface) is nearly completely restored at the back
interface, but some parts of y- and z- components, which have
s-polarization are lost during the imaging. The absolute values of
electrical fields in source and image planes are plotted in Figs.
\ref{m98}.G,H for comparison with Figs. \ref{num}.C,D. The probe
used for the measurements does not allow to determine exact value of
the local field: it actually provides an averaged value by a volume
about $1{\rm cm}\times 1{\rm cm}\times 1{\rm cm}$ around the probe.
That is why the maxima visible at Fig. \ref{num}.D related to the
terminations of the wires can not be observed at Fig. \ref{m98}.H.
In spite of that the general behaviour of the measured field and
distribution predicted by numerical simulations are identical.

Further experimental studies show that the imaging with 2cm
resolution observed at 980MHz keeps within 920MHz--1.1GHz frequency
range. It means the the bandwidth of the wire medium lens is about
18\% which is incomparably wider than bandwidth of operation for the
Swiss rolls. The similar imaging effects are also experimentally
observed within 1.9--2.1 GHz frequency range. It corresponds to the
Fabry-Perot resonance appearing when the thickness of the slab is
equal to the whole wavelength. This confirms that the imaging exists
when the thickness of slab becomes equal to integer number of
half-wavelengths and observed effects are well described by the
canalization regime. Therefore, the lens can be made arbitrary thick
and no distortions will appear. The ohmic losses caused by the
currents in the wires do not disturb the imaging quality. These
losses are only able to reduce intensity of the image, in the same
manner as it happens in long multi-conductor transmission lines.

In conclusion, it is necessary to note that in the present paper we
proposed the lens formed by conducting wires capable to transmit
images with sub-wavelength resolution for long distances as compared
to the wavelength. The lens operates in the canalization regime when
any incident spatial harmonics transforms inside the crystal into
the plane waves which deliver image from one interface to another.
The plane waves in the crystal travel with the same phase velocity
which allows to tune thickness of the slab to fulfill Fabry-Perot
condition for any incident angle and achieve total transmission.
This imaging effect was verified both numerically and experimentally
at gigahertz frequencies. The sub-wavelength resolution of
$\lambda/15$ and 18\% bandwidth of operation are demonstrated. It is
shown that the system is not sensitive to losses in the wires. All
listed effects become possible only due to presence of transmission
line modes in wire medium and strong spatial dispersion
\cite{WMPRB}. The wire medium effectively becomes anisotropic
dielectric with infinite permittivity along anisotropy axis and it
is caused not by frequency resonance as in the case of Swiss rolls
but by strong spatial dispersion effect. Following \cite{WMPRB}, the
component of wire medium permittivity tensor corresponding to the
direction along the wires in the spectral region has the form:
$$
\varepsilon_\parallel(\omega,q)=\varepsilon_0\left(1-\frac{k_0^2}{k^2-q^2}\right),
$$
where $\omega$ is frequency, $q$ is component of wave vector along
wires, $k=\omega/c$ is wavenumber, $c$ is the speed of light and
$k_0$ is wavenumber corresponding to the plasma frequency. The
transmission line modes travel along wires and have $q=k$. It means,
that for such modes the wire medium effectively has infinite
permittivity. The losses in wires influence to the plasma frequency
making it complex \cite{StasMOTL} and thus do not affect the value
of permittivity corresponding to transmission line modes.

The present realization of canalization regime with the help of wire
medium is advantageous at microwave frequencies, but it can not be
realized at optical range where metals lose their conducting
properties. It does not mean that implementation of canalization
regime in optical range is impossible. This regime can be realized
using photonic crystals \cite{Allanglediag,Parimilens,canal}, but
resolution of such lenses will be restricted by period of the
crystals which can not be reduced too much due to absence of
high-contrast lossless materials at the optical range. The other
possibility is to construct uniaxial material with infinite
permittivity along anisotropy axis. It can be done using lattices of
resonant uniaxial nanoparticles or multilayered structures
\cite{bundle1,bundle2}.

The authors acknowledge Mr. John Dupuy for the construction of the
lens. P.A.B. would like to thank Dr. Stanislav Maslovski who
originally brought to his attention the idea of the image transfer
using wire medium in 2002, and Profs. Constantin Simovski and Sergei
Tretyakov for useful discussions.

\bibliography{wml}
\end{document}